# Selective Photothermal Eradication of Glioblastoma Cells Coexisting with Astrocytes by anti-EGFR Coated Raman Tags

YUNG-CHING CHANG[1], CHAN-CHUAN LIU[2,3,4], WAN-PING CHAN[1], YU-LONG LIN[1], CHUN-I SZE[2,4], SHIUAN-YEH CHEN[1*]

[1] Dept. of Photonics, National Cheng Kung University, Tainan, Taiwan

[2] Dept. of Cell Biology and Anatomy, National Cheng Kung University, Tainan, Taiwan

[3] National Institute of Cancer Research, National Health Research Institutes, Tainan, Taiwan

[4] Institute of Basic Medical Sciences, College of Medicine, National Cheng Kung University, Tainan, Taiwan.

Abstract: Glioblastoma (GBM) is an aggressive and fatal tumor. The infiltrative spread of GBM cells hinders the gross total resection. The residual GBM cells are significantly associated with survival and recurrence. Therefore, a theranostic method that can enhance the contrast between residual GBM and normal astrocyte (AS) cells as well as selectively eradicate GBM cells is highly desired. In this report, GBM and normal astrocyte cells are both cultured in the same microplate well to imitate a coexistence environment and treated with Raman tags functionalized by anti-EGFR. Compared to AS cells, GBM cells show 25% higher Raman emission, and their cell death rate increases by a factor of 2. These results demonstrate potential for selective eradication of the residual GBM cells guided by robust Raman signals after the primary GBM surgery.

1. Introduction

Glioblastoma (GBM) is the most prevalent malignant brain tumor in adults, constituting 45-50% of all primary malignant brain neoplasms [1]. According to the World Health Organization's (WHO) classification criteria, GBM is classified as a central nervous system grade 4 tumor [1]. The median survival of the patients treated by the standard treatment is only 14.6 months [2], and the five-year survival rate is 9.8% [3]. The standard treatment for individuals newly diagnosed with GBM involves undergoing neurosurgery to remove the tumor mass, followed by concurrent radiation therapy and daily doses of temozolomide. Afterward, they typically undergo six cycles of temozolomide [2]. GBM is notorious for its infiltrative growth, which hinders precise resection. Infiltrating GBM cells, which escape initial surgical resection and other initial therapies, are the most probable source of local recurrence [1]. Since the extent of resection and residual volume are associated with survival and recurrence [4], for first-line surgery, the method to enhance the contrast between GBM tumors and surrounding normal tissue as well as efficiently eliminate the GBM tumor cells is highly desired. In 2017, the US FDA approved fluorescence-guided surgery (FGS) for high-grade gliomas [5, 6]. Patients treated with FGS showed improvement in completeness of tumor resection compared to conventional white light (65% vs. 36%) and higher 6-month progression-free survival [5].

However, the mixed evidence shows no improvement in overall survival [5] or some improvement [7] (17 months vs. 10 months) for the patients with FGS. In addition, a recent study showed that 5–ALA–induced fluorescence has low accuracy in classifying the fresh tissue samples into GBM or normal tissue [8]. Therefore, following primary resection, adjuvant therapies to detect and eliminate residual GBM cells remain essential.

In the past decade, exploiting the interaction between nanomaterials and cell/tissue microenvironment has improved drug delivery, diagnosis, target therapy for tumor treatments [9]. Various nanoparticle-mediated treatments for GBM have been proposed and demonstrated [10, 11]. Nanoparticles can serve as imaging, therapeutic and diagnostic agents [12-14]. One of the important therapeutic nanoparticles is based on photothermal therapy (PTT) [15-33]. Those PTT nanoparticles with various blood-brain barrier (BBB) crossing strategies are normally delivered to the tumor site through circulation [21-33]. These strategies include transporter-mediated transcytosis, receptor-mediated transcytosis, cell-mediated transcytosis, liphophilic pathway, efflux pumps, adsorptive transcytosis, paracellular aqueous pathway [34, 35]. On the other hand, the local delivery of the anti-cancer drugs [36] or therapeutic nanoparticles [18] to the resection cavity caused by the first-line GBM surgery can reach residual GBM cells without crossing BBB. However, there are few preclinical efficacy assessments of imaging or therapeutic nanoparticles locally delivered to the tumor resectional cavity where GBM and normal cells coexist. Since EGFR amplification is the most common EGFR alteration in GBM and is observed in 40% of GBM [1], the anti-EGFR functionalized nanoparticles can be selectively attached to the GBM cells. A previous study [37] has used anti-functionalized nano-rods to eradicate two GBM cell lines (U373-MG, 1321N1) and focused on comparing eradication of the cells without nano-rods, with unfunctionalized nano-rods, and with functionalized nano-rods. In this study, we focus on the selective eradication of GBM/normal astrocytes since the coexistence of GBM/normal astrocytes is expected along the margin of the resectional cavity.

In this work, Raman tags based on gold nanoparticle assemblies are constructed to provide a stable Raman and photothermal source. Rat GBM cells (CNS-1) and normal rat astrocytes (AS) are both cultured in the same well to imitate the coexistence of residual GBM cells and surrounding astrocytes at the margin of the cavity caused by the primary resection, Fig. 1. One bare-Tag (b-Tag) and two Raman tags (R-tag1, R-tag2) are assembled and coated with anti-EFGR to selectively bind to GBM cells, Fig. 2(A). The b-Tag is utilized to find the approximate illumination condition for photothermal eradication. R-Tag1 and R-Tag2 are used to acquire the Raman contrast and lead to photothermal eradication simultaneously.

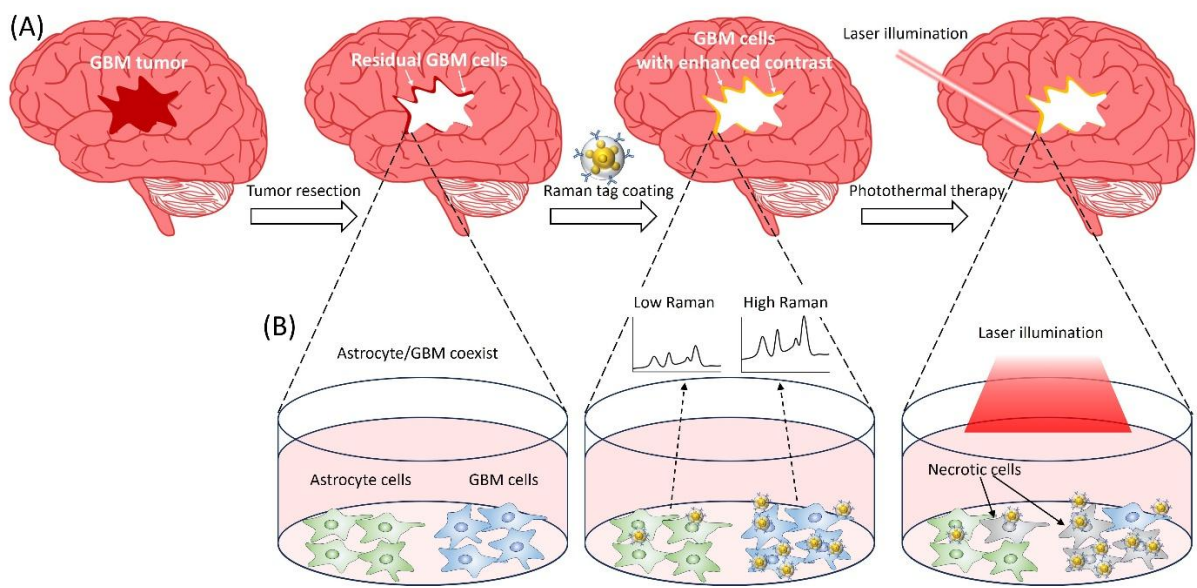

Fig. 1. (A) The schematic illustration of the proposed photothermal eradication of residual GBM cells after the conventional tumor resection. (B) The corresponding in vitro experiments in this study show the GBM cells can be selectively eliminated.

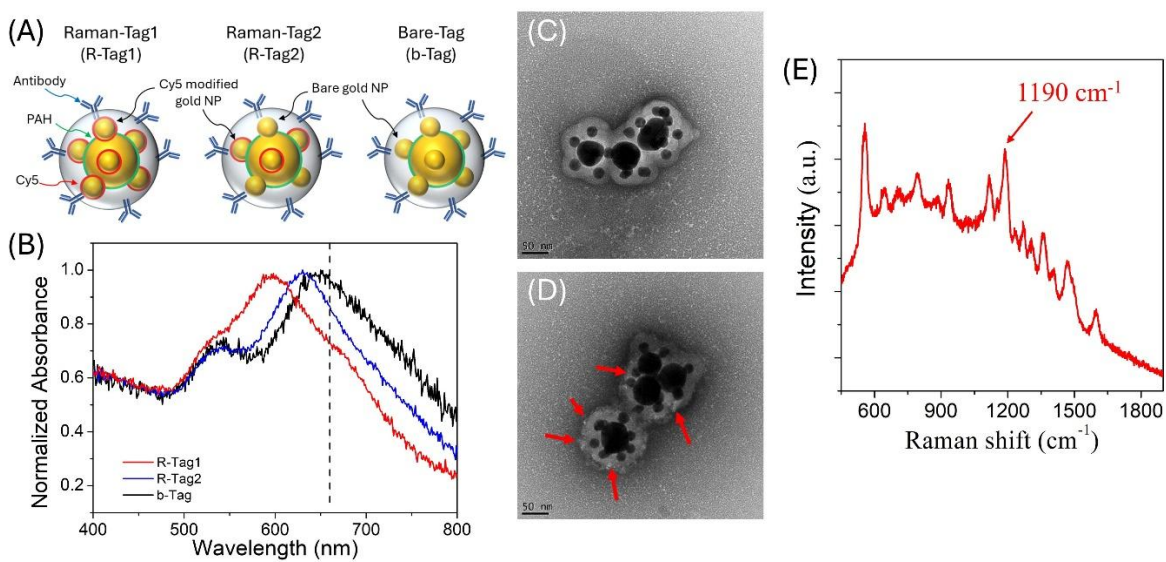

Fig. 2. (A) Illustration of b-Tag, R-Tag1 and R-Tag2. (B) The corresponding extinction spectra. Dashed line: the wavelength of the laser for photothermal eradication. The TEM images of R-Tag2 before (C) and after (D) antibody coating. (E) Raman signals emitted from R-Tag1.

2. Results and Discussion

2.1 Characterization of bare tags and Raman tags

The tags (Fig. 2(A)) are characterized by three methods. The resonance wavelength of the tags is obtained by the extinction spectrum. As shown in Fig. 2(B), R-Tag1, R-Tag2, and b-Tag have primary resonance wavelengths at 598 nm, 631 nm, and 647 nm, respectively. This resonance results from the coupling between core and satellite NP, while the secondary resonance around 540 nm is from the single particle mode. The success of antibody modification is verified by chemical bonds shown in the FTIR spectrum, such as 1541 $cm^{-1}$ and 1644 $cm^{-1}$ for Amide II and Amide I bonds, respectively. The details are described in our previous report [38]. The binding of antibodies can also be confirmed directly by observation of the TEM images before and after the antibody coating. In Fig. 2(C), the surface of the silica shell is smooth, while in Fig. 2(D), it is straightforward to observe negatively stained antibodies (bright spots pointed by red arrows).

The distinct peaks of the predefined Raman feature from Cy5 can be observed from Raman tags, Fig. 2(E). The intensity of the prominent Raman peak at 1190 cm$^{-1}$ is used to calculate the Raman contrast.

2.2 Effectiveness of photothermal eradication by b-Tag

The photo-induced eradication of CNS-1 cells can be achieved by choosing the proper conditions, such as tag concentration, incubation time, irradiance, and illumination time. Once the photo-induced eradication of CNS-1 is observed, we must verify whether this effectiveness results from the photothermal effect or the laser-induced toxicity. The impact of intrinsic laser-induced toxicity is verified with CNS-1 treated with no tags. In addition, if the intrinsic laser intensity has low or no effect on cell death, we must clarify that the photothermal eradication results from specific binding or non-specific binding of the b-tag. Therefore, the CNS-1 cells are also treated with b-Tags coated with anti-EGFR (b-Tag@anti-EGFR) and b-Tags coated with isotype IgG (b-Tag@IgG). The anti-EGFR antibody is a specific antigen-recognizing immunoglobulin G (IgG) while an isotype IgG does not bind the target antigen. Isotype IgG would bind to irrelevant antigen due to non-specific interactions, such as sticky surfaces, solution condition, and other experimental artifacts.

For clarity, each of the following experiments is labeled as Exp#, and their complete experimental conditions are listed in Table S1. In Exp1, the following parameters are used: incubation times (4.5 hr, 9 hr, 17 hr), tag concentration (4.3 pM), and irradiance 820W/cm$^2$). As shown in Fig. 3(A) and Fig. S3, the laser illumination with b-Tag@anti-EGFR, b-Tag@IgG, or no tags does not lead to cell necrosis, no matter which incubation time is adopted.

Then, in Exp2 (Table S1), the particle concentration is increased to 8.6 pM, while the laser power density is decreased to 410 W/cm$^2$. As shown in Fig. 3(B) and Fig. S4, cell necrosis is only observed from cells labeled with b-Tag@anti-EGFR. The average cell death rate is around ~50%, while for CNS-1 treated with b-Tag@IgG and no tags, the average death rate is <2%. The low death rate for cells treated with b-Tag@IgG proves non-specific binding cannot cause sufficient photothermal eradication in Exp2. In addition, for the group without tags, the low death rate indicates that the laser induced toxicity is insignificant. Only CNS-1 treated with b-Tag@anti-EGFR can cause significant photothermal eradication of GBM cells.

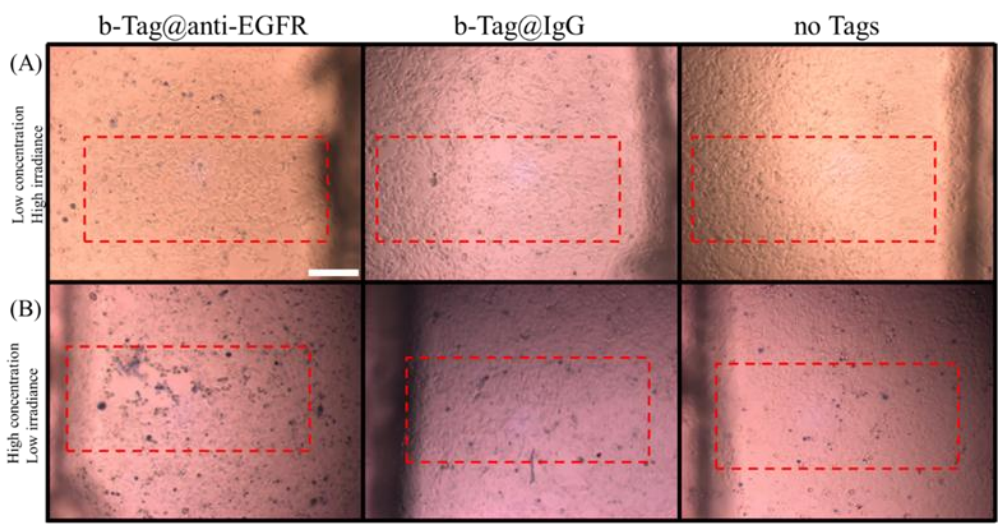

Fig. 3. Representative images of CNS-1 cells after photothermal illumination. The laser scanning region is highlighted by the red rectangle. Scale bar: 200 um. The incubation time is 9 hr. (A) Exp1: low tag concentration (4.3 pM)/ high irradiance (820 W/cm$^2$). (B) Exp2: high tag concentration (8.6 pM)/low irradiance (410 W/cm$^2$).

2.3 Selectivity of photothermal eradication by b-Tag for CNS-1 and AS

Since Exp2 (Table S1) shows the photothermal eradication of CNS-1 cells with b-Tag@anti-EGFR, in Exp3 (Table S1), we apply this condition to see if it can selectively eradicate CNS-1 rather than AS cells. In addition, three lengths of illumination time are also adopted to investigate if it will affect the outcome of photothermal eradication as shown in Fig. 4(A) (complete cell images in Fig. S5). For CNS-1, the death rates of three illumination times are similar (~20%), while for AS, the death rate is increased with a longer illumination time. Thus, the contrast of the death rate is not apparent for 3 and 5 min. illumination but more obvious for 1 min. illumination. However, it is not conclusive due to the variation of cell death rate.

In Exp4 (Table S1), the concentration is further increased to 25 pM, and the incubation time is decreased to 3 hr. As shown in Fig. 4(B) (original images in Fig. S6), in all three illumination lengths, the death rates of CNS-1 (24.1%, 28.6%, 20.4%) are all higher than AS (9.2%, 4.6%, 11.4%). Among them, 1 and 3 min. illuminations have higher cell death rate contrast than the 5 min. illumination. Although the cell death rate contrast is improved in Exp4, statistical significance for the cell death rate of CNS-1/AS has not yet been achieved.

So far, the approximate conditions for selectively eradicating CNS-1/AS seem obtainable. In Exp5 (Table S1), CNS-1 and AS are co-cultured in the same well to imitate the coexistence environment at the margin of a resection cavity. The b-Tag concentration is slightly increased to 33 pM. The significant death rates contrast is shown in Fig. 4(C) (original images in Fig. S7). Under 1 min. illumination, the death rate is 60% in CNS-1, higher than the 19% in AS. Under 3 min. illumination, the death rate is 43% in CNS-1, also higher than the 13% in AS. Thus, the conditions in Exp5 can selectively eliminate CNS-1 labeled by b-Tag.

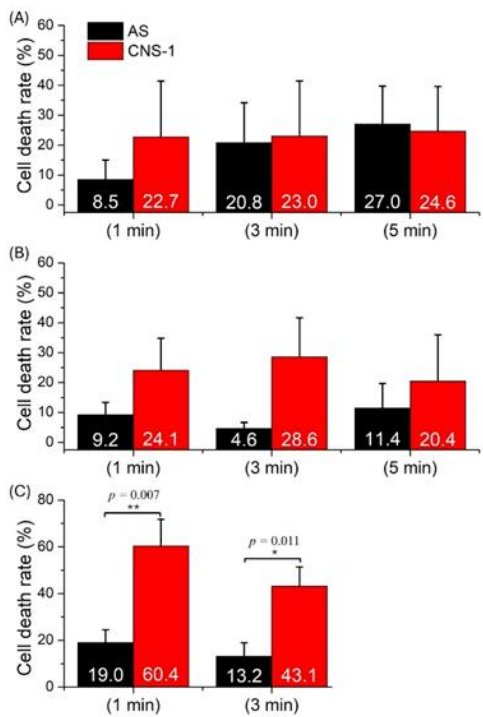

Fig. 4 The cell death rate of AS and CNS-1 treated with b-Tag vs. illumination time under different tag concentrations and incubation times. (A) 8.6 pM, 9 hr. (B) 25 pM, 3 hr. (C) 33 pM, 3 hr.

2.4 Photothermal eradication and Raman contrast of CNS-1/RA by R-Tag1

Since the effectiveness and selectivity of the photothermal eradication by b-Tag are achieved in Exp5, the Raman tag (R-Tag1) replaces b-Tag in Exp6 to add Raman contrast to CNS-1/AS cells. Observation of both Raman contrast and selective photothermal eradication is expected. The condition in Exp6 (Table S1) is the same as Exp5 except that R-Tag1 is used, and 5 min. illumination is included again. However, compared to the cells treated with b-Tag in Fig. 4, CNS-1 and AS cells labeled with R-Tag1 scarcely show cell necrosis (data not shown).

Therefore, in Exp7 (Table S1), we increase the reaction time from 3 hr to 4/8 hr and measure the Raman signals first. The Raman signal is always more prominent in CNS-1, which means that selective adsorption of R-Tag1 occurs. For the 4 hr incubation, the average Raman contrast between CNS-1/AS is 1.46; for the 8-hour incubation, the Raman contrast decreases to 1.2. The lower Raman contrast shown in the group with longer incubation (8 hr) may result from the adsorption of R-Tag1 due to the non-specific binding of anti-EGFR to the AS cells. The longer incubation time increases the chance of non-specific binding. For all 18 observation points, cell necrosis is only observed in two points (data not shown). The prominent Raman contrast indicates the selective adsorption of R-Tag remains effective. However, the failure of the photothermal eradication implies that the photothermal efficiency of R-tag1 may be substantially lower than b-Tag. In addition, the cytotoxicity of Raman tags to GBM and AS cells are also performed in Exp7. Under 4/8 hrs incubation with Raman tags but without laser illumination, no obvious cell death is observed which indicates the Raman tags are non-toxic to GBM and AS cells, as shown in Fig. S8.

Under the same tag concentration and illumination time, R-Tag1 fails to reproduce the effectiveness and selectivity of photothermal eradication shown by b-Tag. This failure may result from the considerable blueshift of the resonance frequency from 647 nm to 598 nm, away from the wavelength of photothermal excitation (660 nm), Fig. 2(B). For R-tag1, the insertion of the Raman reporter, oligonucleotide-modified Cy5, enlarges the gap between core and satellite nanoparticles and reduces the plasmonic coupling. Thus, the blueshift reflects the reduced coupling. The extinction of R-Tag1 is significantly lower at 660 nm, which may lead to inefficiency of photothermal conversion. Therefore, the efficacy of photothermal eradication by R-Tag1 dramatically deteriorates.

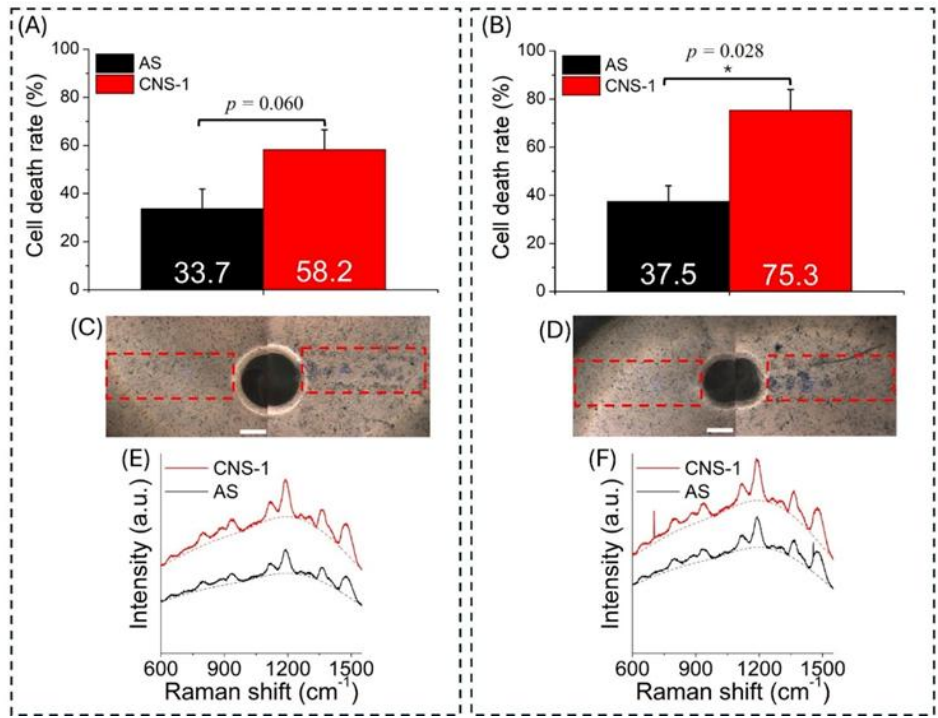

Fig. 5. CNS-1/AS cells treated with R-tag2 under (Left) 1 min. and (Right) 3 min. photothermal illumination. CNS-1/AS death rate: (A)(B). Representative cell images: (C)(D) (scale bar: 200 um) and the corresponding Raman spectra: (E)(F).

2.5 Photothermal eradication and Raman contrast of CNS-1/RA by R-Tag2

To recover the photothermal efficiency of Raman tags, the redshift of resonance close to 660 nm is necessary. Therefore, half of the Cy5 functionalized satellite NP are replaced by the bare satellite NP to construct R-Tag2. As shown in Fig. 2(B), the peak resonance of the extinction of the R-Tag2 is substantially redshifted to 631 nm.

In Exp8 (Table S1), R-Tag2@anti-EGFR is applied to the CNS-1/AS coexistence environment to observe the cell death rate and Raman contrast. Among nine illuminated regions (6 for 1 min illumination, 3 for 3 min illumination), CNS-1 cells have a higher death rate and Raman signal strength than AS in 8 regions.

Under 1 min illumination, the average death rate is 33.7% and 58.2% for AS and CNS-1, respectively, Fig. 5(A)(C). The average death rate contrast is 2.12, and the average Raman contrast is 1.26, Fig. 5(E). The correlation coefficient for death rate contrast and Raman contrast is 0.88. Under 3-minute illumination, the average death rate is 37.5% and 75.3% for AS and CNS-1, respectively, Fig. 5(B)(D). The average death rate contrast is 2.21, and the average Raman contrast is 1.23, Fig. 5(F). The correlation coefficient for death rate contrast and Raman contrast is 0.52.

Interestingly, a similar average death rate contrast is observed under both photothermal illumination lengths (2.12 vs. 2.21). The photothermal effect causes a substantially higher death rate in CNS-1 by a factor of 2. Thus, CNS-1/AS cells labeled with R-Tag2 demonstrate the Raman contrast and selective photothermal eradication in the coexistence environment.

2.6 Discussion

The tag concentration and resonance wavelength of the tags are the two critical factors of the effectiveness of photothermal eradication. According to Exp1 and Exp2, the higher concentration leads to the initial efficacy of the photothermal eradication of CNS-1 cells. For the effect of illumination time on GBM cells, under lower tags concentration (8.6 pM in Fig. 4(A) and 25 pM in Fig. 4(B)), the GBM death rate is not sensitive to the illumination time and keeps ~25%. Under higher tag concentration (33 pM in Fig. 4(C) and 43 pM in Fig. 5), the GBM cell death significantly increases to >40% but is still not sensitive to the illumination time. Furthermore, in Exp8, the similar death rate contrast is observed for 1 min. and 3 min. illumination. Therefore, compared to tag concentration and resonance wavelength, illumination time is probably not the key factor under our experimental conditions.

Exp3 to Exp5 also show that the increasing tag concentration results in better selectivity of photothermal eradication of CNS-1/AS cells. Comparing b-Tag (Exp5) to R-Tag1(Exp6) indicates that the offset of the resonance wavelength is detrimental to the photothermal performance of the tags. The R-tag2 in Exp8 demonstrates the selective photothermal eradication of CNS-1 again, even though its extinction is slightly lower than the b-Tag. The b-Tag in Exp5 (Fig. 4(C)) has better selectivity of photothermal eradication than R-Tag2 in Exp8 (Fig.5(A)(B)). However, the R-tag2 can provide Raman and cell death rate contrast simultaneously.

A thinner silica shell could enhance the photothermal conduction of Raman tags. The current 20 nm thickness could be decreased to <5 nm by the other silica coating process such as silicate deposition. The ~25% Raman contrast can be further enhanced by increasing the brightness of the Raman tags through tuning the resonance of the assembly. Increasing the concentration of the Raman tags might also improve the Raman contrast.

### 3. Conclusion

This work demonstrates the selective photothermal eradication of GBM (CNS-1)/astrocyte (AS) cells when cells are cultured in the same well and treated with Raman tags. The cell death rate of CNS-1 is higher than that of AS by a factor of 2. In addition, the Raman signal of the CNS-1 region is 25% higher than AS cells. Combining Raman contrast and selective photothermal eradication, this work demonstrates the possibility of eliminating the residual GBM cells around the edge of the resection cavity after primary surgery. The Raman tags can be further integrated with hydrogels for controlled release. In addition, the portable Raman system can be integrated with a photothermal laser to perform residual GBM cell elimination guided by Raman signals.

### 4. Methods

#### 4.1 Bare and Raman tags preparation

In this work, a bare tag (b-Tag) and two Raman tags (R-Tag1, R-Tag2) are all based on a core-satellite assembly (CSA) consisting of a 50 nm gold nanoparticle (NP) as a core surrounded by several 20 nm gold NP as satellites through positively charged polyelectrolytes, Fig. 2(A). The b-Tag has no embedded Raman reporters, so its assembly time is one day less than Raman tags. Thus, b-Tag is used to promptly obtain approximate illumination conditions, including illumination time, power density, tag concentration, etc., for photothermal eradication. The first Raman tag, R-Tag1, has a core particle surrounded by the 20 nm gold NP coated with oligonucleotide-modified Cy5. Cy5 is the primary Raman reporter, Fig. 2(E). Oligonucleotide is used to keep the surface charge of the gold NP negative. The second Raman tag, R-Tag2, has the same configuration as R-Tag1 except that the bare 20 nm NP replaces half of the Cy5-coated 20 nm NP to tune the resonance frequency close to laser excitation to optimize the photothermal effect. The core-satellite assembly is coated with a silica shell and then functionalized with anti-EGFR to selectively bind to GBM cells. The isotype IgG is functionalized for the control group.

To construct b-Tag, 800 uL of citrate-stabilized 50 nm gold nanoparticle suspension (BBI) is diluted by 720 uL DI water, then mixed with 80 uL of 6 mM polyallylamine hydrochloride (PAH). The mixed solution is centrifuged at 2040 g for 15 min three times to remove excess PAH. The supernatant is removed for the first two centrifugations, and the same amount of DI water is added. For the third centrifugation, after removing the supernatant, DI water is added to reach the total volume of 435 uL. This PAH-coated gold NP suspension is extracted by a 1 mL syringe. The 617 uL of 20 nm gold NP suspension (BBI) is prepared in the other 1 mL syringe. The needles of two syringes are separated by ~2 mm. The 20 nm and 50 nm gold NP suspensions are ejected from the tip of the syringe needles and then mixed drop by drop. The total mixing time is around 5 min, and the mixed suspension's color shifts from red to blue due to the plasmonic resonance coupling between the core and satellite gold NP.

To construct the R-Tag1, first, citrate-stabilized 20 nm gold nanoparticles (1.16 nM, BBI solution) are mixed with oligonucleotide-modified Cy5 through the salt aging method. 125 uL of oligonucleotide-modified Cy5 (2.95 uM) is added to 800 uL of 20 nm AuNP. 1 M of NaCl and 0.1 M of sodium phosphate buffer (pH=7) are prepared with a volume equal to one-eighth of the mixture of Gold NP/Cy5-modified oligonucleotide. The ⅛ aliquot of NaCl and sodium phosphate buffer is respectively added every one hour for the first eight hours. After 16 hr of incubation, the solution of Cy5-modified 20 nm gold NP suspension is centrifuged at 5200 g for 20 min three times. After the third removal of the supernatant, DI water is carefully added to maintain the initial concentration of 20 nm gold NP. The 435 uL Cy5-modified 20 nm NP and 617 uL of 50 nm PAH-coated gold NP are mixed to form R-Tag1.

The assembly of R-Tag2 is the same as R-Tag1 except that after the last centrifugation of 20 nm NP, the DI water is refilled with DI water to a higher concentration of 2.8 nM. The 126.5 uL of Cy5-modified 20 nm NP and 308.5 uL of citrate stabilized 20 nm NP are mixed. The 435 uL of 20 nm NP is mixed with the same amount of PAH-modified 50 nm NP through the needle tip of the syringe. After being left to stand for 15 min, 182 uL of DI water is added to obtain 1052 uL of R-Tag2 suspension.

All three tags are coated with a silica shell followed by anti-EGFR. Here, the processes are described. The detailed steps can be found in our previous report [38, 39]. For silica coating, APTMS ((3-Aminopropyl)trimethoxysilane, Alfa Aesar) is added to produce an anchor layer where silica would grow, followed by adding TEOS (tetraethyl orthosilicate, Showa). Initial silica shell is formed by hydrolysis of TEOS. Then, APTMS is added again to increase the formation of the amino group on the surface of silica shell. The whole solution is incubated at 35°C and 800 rpm for 24 hr. For anti-EGFR functionalization, the amino group (-NH$_2$) is replaced by carboxyl group (-COOH) by SA (Succinic anhydride, Acros Organics). EDC (N-Ethyl-N'-(3- dimethylaminopropyl)carbodiimide hydrochloride, Sigma-Aldrich) and sulfo-NHS (NHydroxysulfosuccinimide sodium salt, Sigma-Aldrich) are added to produce active ester bond. Anti-EGFR can react with the ester bond to form amide bond. After the functionalization of anti-EFGR, 0.25% of BSA is added to the tag suspension to increase stability.

4.2 Cell preparation

The rat GBM (CNS-1) and rat astrocyte (AS) cells are grown in 3.5 cm culture dishes. The subculture is performed when the population of cells reaches 70% for CNS-1 and 90% for AS. To seed cells in a 96-well microplate, the cells in the culture dish are suspended by Trypsin, counted with cell counters, and then diluted with Dulbecco's modified eagle medium (DMEM) to the working concentration: $5 \times 10^4$/mL for CNS-1 and $11.5 \times 10^4$/mL for AS. The 200 uL of cell suspension is injected into a well of the microplate. Due to the higher cell growth rate of CNS-1, cell expansion time is 19 hr for CNS-1 and 24 hr for AS.

4.3 GBM and AS Cells prepared in the same microplate well

120 uL of DMEM is added into a well of a 96-well microplate. The PP film coated with PDMS along the edge is inserted into the middle of the well to divide the well into two compartments. 15 uL of AS cell suspension is added into a compartment of the well. Five hours later, 15 uL of CNS-1 suspension is added into the other compartment. After 19 hours of incubation, the PP spacer is removed, and then CNS-1 and AS cells are exposed to the same external condition. For each well, 110 uL of DMEM is extracted, and then 90 uL of Raman tags suspension is added. The incubation time ranges from 3 to 17 hr, depending on the experiment conditions. After the incubation, 50 uL of the medium is removed. The whole sample is rinsed with 200 uL of 1x PBS three times. Then, 200 uL of serum-free medium is added to the well for 60 min incubation. Finally, CNS-1 and AS are ready for Raman acquisition and photothermal illumination in this coexistence environment.

4.4 Raman signals acquisition

Raman signals from Raman tags attached to CNS-1 and AS cells in the coexistence environment are acquired by an inverted microscope with a 632.8 nm HeNe laser and a spectrometer (grating is 1200 l/mm with slit size 300 um). The laser is focused by a 10x objective lens, concentrating the laser light into a point ~200 um in diameter. A marker is engraved at the backside of the center of the target well. The marker is used as a positional reference point. For each side of the cells, the Raman signal is acquired sequentially from 5 points equally spaced (250 um) along a straight line. After Raman acquisition, the 96-well microplate is moved back to the incubator for 30 min to stabilize the cells.

4.5 Photothermal setup and procedure

After the Raman acquisition, the whole microplate is moved to a custom photothermal setup. The 660 nm CW laser beam is focused to a spot of ~200 um in diameter. This work adopts two irradiance settings (410 and 820 W/cm$^2$). Three illumination times (1 min, 3 min, and 5 min) for each illuminated spot are utilized. The spacing of each spot is 150 um.

4.6 Cell death rate calculation

To evaluate the efficacy of photothermal treatment, the cell death rate of the illuminated cells must be defined. After the photothermal treatment, Trypan blue staining is performed to determine cell death. The dead cells become blue. The whole blue area is integrated by Image J and then divided by the illuminated area to obtain the cell death rate.

4.7 TEM characterization of anti-EGFR functionalized Raman tags

Anti-EGFR functionalized Raman tags are characterized by TEM. First, the sample is processed by negative staining. The carbon film of a TEM grid (CF200-Cu, EMS) is hydrophilized by a hydrophilizing system (JEOL HDT-400). The 4 uL of tag suspension is deposited on the TEM grid. After 50 s standing, the excess suspension is removed by filter paper. The grid is rinsed with 4 uL of DDI water and then the excess water is removed. Then, 0.1 M of uranyl acetate is dropped onto the grid, kept for 50s and then removed by filter paper. After the grid is dried, the sample is ready for observation under TEM (JEOL, JEM-1400).

**Supporting Information**. S1: Experimental conditions for photothermal eradication experiments from Exp1 to Exp8.  S2: The photothermal profiles and photothermal images of the tags. S3-S7: The complete set of cell images acquired after photothermal eradication in Exp1 to Exp5. S8: The cytotoxicity test of Raman tags in Exp7.  S9: The complete set of cell images acquired after photothermal eradication in Exp8.

AUTHOR INFORMATION

**Corresponding Author**

Shiuan-Yeh Chen

*sychen72@ncku.edu.tw

**Funding Sources**

National Science and Technology Council, Taiwan (MOST 106-2221-E-006-171, MOST 107-2221-E-006-147, NSTC 111-2221-E-006-063-MY3).

**Notes**

The authors declare no conflicts of interest.

REFERENCES

(1) WHO Classification of Tumours Editorial Board. *Central Nervous System Tumours*; Medicine Series, Vol. 6; International Agency for Research on Cancer, 2022. ISBN : 9789283245087
(2) Stupp, R.; Weller, M.; Belanger, K.; Bogdahn, U.; Ludwin, S. K.; Lacombe, D.; Mirimanoff, R. O. Radiotherapy plus Concomitant and Adjuvant Temozolomide for Glioblastoma. *The New England Journal of Medicine* **2005**, *352* (10), 987–996.
(3) Stupp, R.; Hegi, M. E.; Mason, W. P. Effects of Radiotherapy with Concomitant and Adjuvant Temozolomide versus Radiotherapy Alone on Survival in Glioblastoma in a Randomised Phase III Study: 5-Year Analysis of the EORTC-NCIC Trial. The *Lancet Oncology* **2009**, *10* (5), 459–466.
(4) Chaichana, K. L.; Jusue-Torres, I.; Navarro-Ramirez, R.; Raza, S. M.; Pascual-Gallego, M.; Ibrahim, A.; Hernandez-Hermann, M.; Gomez, L.; Ye, X.; Weingart, J. D.; Olivi, A.; Blakeley, J.; Gallia, G. L.; Lim, M.; Brem, H.; Quinones-Hinojosa, A. Establishing Percent Resection and Residual Volume Thresholds Affecting Survival and Recurrence for Patients with Newly Diagnosed Intracranial Glioblastoma. *Neuro-Oncology* **2014**, *16* (1), 113–122. https://doi.org/10.1093/neuonc/not137.
(5) Stummer, W.; Pichlmeier, U.; Meinel, T.; Wiestler, O. D.; Zanella, F.; Reulen, H.-J. Fluorescence-Guided Surgery with 5-Aminolevulinic Acid for Resection of Malignant Glioma: A Randomised Controlled Multicentre Phase III Trial. *The Lancet Oncology* **2006**, *7* (5), 392–401. https://doi.org/10.1016/S1470-2045(06)70665-9.
(6) Hadjipanayis, C. G.; Stummer, W. 5-ALA and FDA Approval for Glioma Surgery. *Journal of neuro-oncology* **2019**, *141*, 479–486.
(7) Baig Mirza, A.; Christodoulides, I.; Lavrador, J. P.; Giamouriadis, A.; Vastani, A.; Boardman, T.; Ahmed, R.; Norman, I.; Murphy, C.; Devi, S.; Vergani, F.; Gullan, R.; Bhangoo, R.; Ashkan, K. 5-Aminolevulinic Acid-Guided Resection Improves the Overall Survival of Patients with Glioblastoma—a Comparative Cohort Study of 343 Patients. *Neuro-Oncology Advances* **2021**, *3* (1), vdab047. https://doi.org/10.1093/noajnl/vdab047.
(8) Livermore, L. J.; Isabelle, M.; Bell, I. M.; Edgar, O.; Voets, N. L.; Stacey, R.; Ansorge, O.; Vallance, C.; Plaha, P. Raman Spectroscopy to Differentiate between Fresh Tissue Samples of Glioma and Normal Brain: A Comparison with 5-ALA–Induced Fluorescence-Guided Surgery. *Journal of Neurosurgery* **2021**, *135* (2), 469–479. https://doi.org/10.3171/2020.5.JNS20376.
(9) Chu, Z.; Wang, W.; Zheng, W.; Fu, W.; Wang, Y.; Wang, H.; Qian, H. Biomaterials with Cancer Cell-Specific Cytotoxicity: Challenges and Perspectives. *Chem. Soc. Rev.* 2024, 53 (17), 8847–8877. https://doi.org/10.1039/D4CS00636D.

(10) Song, X.; Qian, H.; Yu, Y. Nanoparticles Mediated the Diagnosis and Therapy of Glioblastoma: Bypass or Cross the Blood–Brain Barrier. *Small* **2023**, *19* (45), 2302613. https://doi.org/10.1002/smll.202302613.

(11) Ghosh, A. K.; Ghosh, A.; Das, P. K. Nanotechnology Meets Stem Cell Therapy for Treating Glioblastomas: A Review. *ACS Appl. Nano Mater.* **2024**, *7* (3), 2430–2460. https://doi.org/10.1021/acsanm.3c04714.

(12) Kircher, M. F.; De La Zerda, A.; Jokerst, J. V.; Zavaleta, C. L.; Kempen, P. J.; Mittra, E.; Pitter, K.; Huang, R.; Campos, C.; Habte, F.; Sinclair, R.; Brennan, C. W.; Mellinghoff, I. K.; Holland, E. C.; Gambhir, S. S. A Brain Tumor Molecular Imaging Strategy Using a New Triple-Modality MRI-Photoacoustic-Raman Nanoparticle. *Nat Med* **2012**, *18* (5), 829–834. https://doi.org/10.1038/nm.2721.

(13) Pal, S.; Ray, A.; Andreou, C.; Zhou, Y.; Rakshit, T.; Wlodarczyk, M.; Maeda, M.; Toledo-Crow, R.; Berisha, N.; Yang, J.; Hsu, H.-T.; Oseledchyk, A.; Mondal, J.; Zou, S.; Kircher, M. F. DNA-Enabled Rational Design of Fluorescence-Raman Bimodal Nanoprobes for Cancer Imaging and Therapy. *Nat Commun* **2019**, *10* (1), 1926. https://doi.org/10.1038/s41467-019-09173-2.

(14) Premachandran, S.; Haldavnekar, R.; Ganesh, S.; Das, S.; Venkatakrishnan, K.; Tan, B. Self-Functionalized Superlattice Nanosensor Enables Glioblastoma Diagnosis Using Liquid Biopsy. *ACS Nano* **2023**, *17* (20), 19832–19852. https://doi.org/10.1021/acsnano.3c04118.

(15) Capart, A.; Metwally, K.; Bastiancich, C.; Da Silva, A. Multiphysical Numerical Study of Photothermal Therapy of Glioblastoma with Photoacoustic Temperature Monitoring in a Mouse Head. *Biomed. Opt. Express* **2022**, *13* (3), 1202. https://doi.org/10.1364/BOE.444193.

(16) Arami, H.; Kananian, S.; Khalifehzadeh, L.; Patel, C. B.; Chang, E.; Tanabe, Y.; Zeng, Y.; Madsen, S. J.; Mandella, M. J.; Natarajan, A.; Peterson, E. E.; Sinclair, R.; Poon, A. S. Y.; Gambhir, S. S. Remotely Controlled Near-Infrared-Triggered Photothermal Treatment of Brain Tumours in Freely Behaving Mice Using Gold Nanostar s. *Nat. Nanotechnol.* **2022**, *17* (9), 1015–1022. https://doi.org/10.1038/s41565-022-01189-y.

(17) De La Encarnación, C.; Jungwirth, F.; Vila-Liarte, D.; Renero-Lecuna, C.; Kavak, S.; Orue, I.; Wilhelm, C.; Bals, S.; Henriksen-Lacey, M.; Jimenez De Aberasturi, D.; Liz-Marzán, L. M. Hybrid Core–Shell Nanoparticles for Cell-Specific Magnetic Separation and Photothermal Heating. *J. Mater. Chem. B* **2023**, *11* (24), 5574–5585. https://doi.org/10.1039/D3TB00397C.

(18) Nie, D.; Ling, Y.; Lv, W.; Liu, Q.; Deng, S.; Shi, J.; Yang, J.; Yang, Y.; Ouyang, S.; Huang, Y.; Wang, Y.; Huang, R.; Shi, W. *In Situ* Attached Photothermal Immunomodulation-Enhanced Nanozyme for the Inhibition of Postoperative Malignant Glioma Recurrence. *ACS Nano* **2023**, *17* (14), 13885–13902. https://doi.org/10.1021/acsnano.3c03696.

(19) Yalamandala, B. N.; Chen, Y.-J.; Lin, Y.-H.; Huynh, T. M. H.; Chiang, W.-H.; Chou, T.-C.; Liu, H.-W.; Huang, C.-C.; Lu, Y.-J.; Chiang, C.-S.; Chu, L.-A.; Hu, S.-H. A Self-Cascade Penetrating Brain Tumor Immunotherapy Mediated by Near-Infrared II Cell Membrane-Disrupting Nanoflakes via Detained Dendritic Cells. *ACS Nano* **2024**, *18* (28), 18712–18728. https://doi.org/10.1021/acsnano.4c06183.

(20) Ramirez Henao, J. F.; Boujday, S.; Wilhelm, C.; Bouvet, B.; Tomane, S.; Christodoulou, I.; Sun, D.; Cure, G.; Romdhane, F. B.; Miche, A.; Dolbecq, A.; Mialane, P.; Vallée, A. Chemo–Photothérapeutic Effect of Polyoxometalate-Stabilized Gold Nanostars for Cancer Treatment. *ACS Appl. Nano Mater.* **2024**, *7* (17), 21094–21103. https://doi.org/10.1021/acsanm.4c04277.

(21) Wang, J.; Liu, Y.; Morsch, M.; Lu, Y.; Shangguan, P.; Han, L.; Wang, Z.; Chen, X.; Song, C.; Liu, S.; Shi, B.; Tang, B. Z. Brain-Targeted Aggregation-Induced-Emission Nanoparticles with Near-Infrared Imaging at 1550 Nm Boosts Orthotopic Glioblastoma Theranostics. *Advanced Materials* **2022**, *34* (5), 2106082. https://doi.org/10.1002/adma.202106082.

(22) Song, W.; Zhang, X.; Song, Y.; Fan, K.; Shao, F.; Long, Y.; Gao, Y.; Cai, W.; Lan, X. Enhancing Photothermal Therapy Efficacy by *In Situ* Self-Assembly in Glioma. *ACS Appl. Mater. Interfaces* **2023**, *15* (1), 57–66. https://doi.org/10.1021/acsami.2c14413.

(23) Sun, R.; Liu, M.; Lu, J.; Chu, B.; Yang, Y.; Song, B.; Wang, H.; He, Y. Bacteria Loaded with Glucose Polymer and Photosensitive ICG Silicon-Nanoparticles for Glioblastoma Photothermal Immunotherapy. *Nat Commun* **2022**, *13* (1), 5127. https://doi.org/10.1038/s41467-022-32837-5.

(24) Zhao, F.; Zhang, X.; Bai, F.; Lei, S.; He, G.; Huang, P.; Lin, J. Maximum Emission Peak Over 1500 Nm of Organic Assembly for Blood–Brain Barrier-Crossing NIR-IIb Phototheranostics of Orthotopic Glioblastoma. *Advanced Materials* **2023**, *35* (22), 2208097. https://doi.org/10.1002/adma.202208097.

(25) Kang, D.; Kim, H. S.; Han, S.; Lee, Y.; Kim, Y.-P.; Lee, D. Y.; Lee, J. A Local Water Molecular-Heating Strategy for near-Infrared Long-Lifetime Imaging-Guided Photothermal Therapy of Glioblastoma. *Nat Commun* **2023**, *14* (1), 2755. https://doi.org/10.1038/s41467-023-38451-3.

(26) Yang, M.; Chen, D.; Zhang, L.; Ye, M.; Song, Y.; Xu, J.; Cao, Y.; Liu, Z. Porphyrin-Based Organic Nanoparticles with NIR-IIa Fluorescence for Orthotopic Glioblastoma Theranostics. *ACS Appl. Mater. Interfaces* **2024**, *16* (28), 35925–35935. https://doi.org/10.1021/acsami.4c03012.

(27) Hill, M. L.; Chung, S.-J.; Woo, H.-J.; Park, C. R.; Hadrick, K.; Nafiujjaman, M.; Kumar, P. P. P.; Mwangi, L.; Parikh, R.; Kim, T. Exosome-Coated Prussian Blue Nanoparticles for Specific Targeting and Treatment of Glioblastoma. *ACS Appl. Mater. Interfaces* **2024**, *16* (16), 20286–20301. https://doi.org/10.1021/acsami.4c02364.

(28) Dash, B. S.; Lu, Y.-J.; Chen, J.-P. Enhancing Photothermal/Photodynamic Therapy for Glioblastoma by Tumor Hypoxia Alleviation and Heat Shock Protein Inhibition Using IR820-Conjugated Reduced Graphene Oxide Quantum Dots. *ACS Appl. Mater. Interfaces* **2024**, *16* (11), 13543–13562. https://doi.org/10.1021/acsami.3c19152.

(29) Fan, Q.; Kuang, L.; Wang, B.; Yin, Y.; Dong, Z.; Tian, N.; Wang, J.; Yin, T.; Wang, Y. Multiple Synergistic Effects of the Microglia Membrane-Bionic Nanoplatform on Mediate Tumor Microenvironment Remodeling to Amplify Glioblastoma Immunotherapy. *ACS Nano* **2024**, *18* (22), 14469–14486. https://doi.org/10.1021/acsnano.4c01253.

(30) Rocha, J. V. R.; Krause, R. F.; Ribeiro, C. E.; Oliveira, N. C. D. A.; Ribeiro De Sousa, L.; Leandro Santos, J.; Castro, S. D. M.; Valadares, M. C.; Cunha Xavier Pinto, M.; Pavam, M. V.; Lima, E. M.; Antônio Mendanha, S.; Bakuzis, A. F. Near Infrared Biomimetic Hybrid Magnetic Nanocarrier for MRI-Guided Thermal Therapy. *ACS Appl. Mater. Interfaces* **2024**, acsami.4c03434. https://doi.org/10.1021/acsami.4c03434.

(31) Liu, J.; Cheng, D.; Zhu, A.; Ding, M.; Yu, N.; Li, J. Neutrophil-Targeting Semiconducting Polymer Nanotheranostics for NIR-II Fluorescence Imaging-Guided Photothermal-NO-Immunotherapy of Orthotopic Glioblastoma. *Advanced Science* **2024**, *11* (39), 2406750. https://doi.org/10.1002/advs.202406750.

(32) Zhang, H.; Guan, S.; Wei, T.; Wang, T.; Zhang, J.; You, Y.; Wang, Z.; Dai, Z. Homotypic Membrane-Enhanced Blood–Brain Barrier Crossing and Glioblastoma Targeting for Precise Surgical Resection and Photothermal Therapy. *J. Am. Chem. Soc.* **2023**, *145* (10), 5930–5940. https://doi.org/10.1021/jacs.2c13701.

(33) Fang, X.; Gong, R.; Yang, D.; Li, C.; Zhang, Y.; Wang, Y.; Nie, G.; Li, M.; Peng, X.; Zhang, B. NIR-II Light-Driven Genetically Engineered Exosome Nanocatalysts for Efficient Phototherapy against Glioblastoma. *J. Am. Chem. Soc.* **2024**, *146* (22), 15251–15263. https://doi.org/10.1021/jacs.4c02530.

(34) Wu, D.; Chen, Q.; Chen, X.; Han, F.; Chen, Z.; Wang, Y. The Blood–Brain Barrier: Structure, Regulation and Drug Delivery. *Sig Transduct Target Ther* **2023**, *8* (1), 217. https://doi.org/10.1038/s41392-023-01481-w.

(35) Zha, S.; Liu, H.; Li, H.; Li, H.; Wong, K.-L.; All, A. H. Functionalized Nanomaterials Capable of Crossing the Blood–Brain Barrier. *ACS Nano* **2024**, *18* (3), 1820–1845. https://doi.org/10.1021/acsnano.3c10674.

(36) Bastiancich, C.; Danhier, P.; Préat, V.; Danhier, F. Anticancer Drug-Loaded Hydrogels as Drug Delivery Systems for the Local Treatment of Glioblastoma. *Journal of Controlled Release* **2016**, *243*, 29–42. https://doi.org/10.1016/j.jconrel.2016.09.034.

(37) Fernández-Cabada, T.; Pablo, C. S.-L. D.; Pisarchyk, L.; Serrano-Olmedo, J. J.; Ramos-Gómez, M. Optical Hyperthermia Using Anti-Epidermal Growth Factor Receptor-Conjugated Gold Nanorods to Induce Cell Death in Glioblastoma Cell Lines. *J Nanosci Nanotechnol* **2016**, *16* (7), 7689–7695. https://doi.org/10.1166/jnn.2016.12570.

(38) Huang, L.-C.; Chang, Y.-C.; Wu, Y.-S.; Sun, W.-L.; Liu, C.-C.; Sze, C.-I.; Chen, S.-Y. Glioblastoma Cells Labeled by Robust Raman Tags for Enhancing Imaging Contrast. *Biomed. Opt. Express* **2018**, *9* (5), 2142. https://doi.org/10.1364/BOE.9.002142.

(39) Chang, Y.-C.; Huang, L.-C.; Chuang, S.-Y.; Sun, W.-L.; Lin, T.-H.; Chen, S.-Y. Polyelectrolyte Induced Controlled Assemblies for the Backbone of Robust and Brilliant Raman Tags. *Opt. Express* **2017**, *25* (20), 24767. https://doi.org/10.1364/OE.25.024767.